%% file: main.tex
\documentclass[sigconf, nonacm]{acmart}
\usepackage{geometry}
\usepackage{lmodern}
\usepackage[T1]{fontenc}
\usepackage[utf8]{inputenc}

\usepackage{svg}
\svgpath{{./imgs/}} 

\usepackage{listings}
\usepackage{xcolor}

\AtBeginDocument{%
  }

\usepackage{xcolor}
\usepackage{caption}
\usepackage{subcaption}

\usepackage{algorithm}
\usepackage{algpseudocode}
\usepackage{soul}

\definecolor{lightgray}{gray}{0.95}
\definecolor{highlightleft}{rgb}{0.85,0.95,1.0}
\definecolor{highlightright}{rgb}{1.0,0.9,0.85}
\definecolor{highlightmerge}{rgb}{0.9,1.0,0.9}

\lstdefinestyle{javacode}{
  language=Java,
  basicstyle=\ttfamily\small,
  numbers=left,
  numberstyle=\tiny,
  numbersep=5pt,
  frame=single,
  backgroundcolor=\color{lightgray},
  tabsize=2,
  breaklines=true,
  showstringspaces=false
}

\usepackage{tikz}
\usetikzlibrary{shapes.geometric, arrows, positioning}
\tikzstyle{process} = [rectangle, rounded corners, minimum width=3.8cm, minimum height=1.2cm,text centered, draw=black, fill=gray!10]
\tikzstyle{decision} = [diamond, minimum width=3.5cm, minimum height=1.2cm,text centered, draw=black, fill=gray!10]
\tikzstyle{arrow} = [thick,->,>=stealth]

\usepackage{graphicx}
\usepackage{amsmath}
\usepackage{booktabs}
\usepackage{hyperref}

\newif\iffinal
\finaltrue   

\iffinal
  \usepackage[colorinlistoftodos]{todonotes}
  \renewcommand{\todo}[2][]{}
  \newcommand{\vl}[1]{}
  \newcommand{\rb}[1]{}
  \newcommand{\pb}[1]{}
\else
  \usepackage[colorinlistoftodos]{todonotes}
  \newcommand{\rb}[1]{{\color{blue}[#1]}}
  \newcommand{\pb}[1]{{\color{red}[#1]}}
  \definecolor{darkgreen}{RGB}{0,100,0}
  \definecolor{darkgreen}{HTML}{006600}
  \newcommand{\vl}[1]{{\color{darkgreen}\b{victor}: [#1]}}
\fi

\begin{document}

\newcommand{\CInUFPE}{%
  \affiliation{%
    \institution{Universidade Federal de Pernambuco}
    \city{Recife}
    \country{Brazil}
  }
}

\author{Victor Lira}
\affiliation{%
  \institution{Instituto Federal de Pernambuco}
  \city{Palmares}
  \country{Brazil}
}
\email{vl@cin.ufpe.br}

\author{Paulo Borba}
\CInUFPE
\email{phmb@cin.ufpe.br}

\author{Rodrigo Bonifácio}
\affiliation{%
  \institution{Universidade de Brasília}
  \city{Brasília}
  \country{Brazil}
}
\email{rbonifacio123@gmail.com}

\author{Galileu Santos}
\CInUFPE
\email{gsj@cin.ufpe.br}

\author{Matheus Barbosa}
\CInUFPE
\email{mbo2@cin.ufpe.br}

\renewcommand{\shortauthors}{Lira et al.}

\title{RefFilter: Improving Semantic Conflict Detection via Refactoring-Aware Static Analysis}

\input{sections/abstract}

\maketitle

\input{sections/introduction}

\input{sections/background}
\input{sections/problem-definition}

\input{sections/proposed-method}

\input{sections/experimental-setup}
\input{sections/experimental-results}
\input{sections/discussion}
\input{sections/threats-to-validity}

\input{sections/conclusion}

\nocite{*}

\bibliographystyle{ACM-Reference-Format}
\bibliography{references}

\end{document}

%% file: sections/abstract.tex
\begin{abstract}
Detecting semantic interference remains a challenge in collaborative software development. Recent lightweight static analysis techniques improve efficiency over SDG-based methods, but they still suffer from a high rate of false positives. A key cause of these false positives is the presence of behavior-preserving code refactorings, which current techniques cannot effectively distinguish from changes that impact behavior and can interfere with others. To handle this problem we present \textit{RefFilter}, a refactoring-aware tool for semantic interference detection. It builds on existing static techniques by incorporating automated refactoring detection to improve precision. RefFilter discards behavior-preserving refactorings from reports, reducing false positives while preserving detection coverage. To evaluate effectiveness and scalability, use two datasets: a labeled dataset with 99 scenarios and ground truth, and a novel dataset of 1,087 diverse merge scenarios that we have built. Experimental results show that RefFilter reduces false positives by nearly 32\% on the labeled dataset. While this reduction comes with a non significant increase in false negatives, the overall gain in precision significantly outweighs the minor trade-off in recall. These findings demonstrate that refactoring-aware interference detection is a practical and effective strategy for improving merge support in modern development workflows.
\end{abstract}

%% file: sections/introduction.tex
\section{Introduction}

Collaborative software development relies on frequent code integration~\cite{mens2002}. Although version control systems support automated merging, developers often face the non-trivial task of resolving conflicts~\cite{shen2023, Wardah2020}. Beyond traditional \emph{textual} conflicts, more subtle and potentially harmful issues emerge when concurrent changes interact at the behavioral level~\cite{shen2021}. 
These situations are known as \emph{dynamic semantics} conflicts, or \emph{interference}, which occur when integrating contributions from two different development branches unexpectedly alters the behavior of either branch or the original base program~\cite{ jesus2024lightweight}. Detecting such interference early is essential to prevent regression and reduce integration efforts~\cite{brun2013, Pastore2017, shao2009, joana14it, yang1992, zhang2022}.

To detect interference and avoid these problems, prior work has proposed techniques based on static analysis. System Dependence Graphs (SDGs) based techniques are expressive but computationally expensive~\cite{barrosfilho2017using}. 
More recent lightweight alternatives~\cite{jesus2024lightweight, dejesus2023} have improved scalability, which is essential for broader adoption. However, these lightweight techniques for interference detection remain limited: a significant part of the reported interference corresponds to false positives. A common root cause of interference false positives is behavior-preserving refactoring~\cite{Silva2016, Tsantalis2022, Tsantalis2018, fakhoury2019, peruma2018, lin2019, cedrim2017, iammarino2019}. 
The existing static analysis tools cannot detect that a change is a refactoring, and therefore doesn't cause interference, which only occur due to behavior changes. 
As a result, developers may be frequently alerted to non interfering changes, increasing merge effort. 
Reducing false positives is a critical concern, as developer time spent investigating invalid warnings is often wasted. \pb{Esse restinho do parágrafo pode tirar se precisar de espaço.} This is particularly problematic given that debugging and testing may already account for more than half of total development costs~\cite{Hailpern2001}, bug fixing is frequent and time-consuming~\cite{Aranda2009}, and software projects still suffer from effort and schedule overruns~\cite{Molokken2003, choetkiertikul2015} \rb{seria poss\'{i}vel incluir uma refer\^{e}ncia mais atual aqui?}. \vl{ok, coloquei uma de 2015. tem algumas mais recentes, mas achei essa melhor por ser de ASE.}

To handle this problem, in this paper we introduce \textit{RefFilter}, a refactoring-aware tool for static semantic interference detection. RefFilter builds upon lightweight static analysis by incorporating refactoring detection into the interference detection pipeline. It leverages state-of-the-art tools such as RefactoringMiner~\cite{Tsantalis2022, Tsantalis2018} and ReExtractor+~\cite{reextractorplus} to identify refactorings performed in the development branches. If all the edits involved in a reported interference correspond to refactorings, RefFilter classifies the report as a false positive and discards it before it reaches developers or code integrators. In summary, RefFilter is a cohesive, static-analysis-based interference detection tool that is explicitly aware of refactorings and designed to improve the quality of semantic interference reports.

We evaluate RefFilter using two datasets. The first is a benchmark dataset with 99 merge scenarios and interference ground truth, previously used in related work. The second is a new dataset of 1,087 merge scenarios, which we contribute as part of this paper. Unlike many prior evaluations that rely on small or potentially biased datasets, our evaluation combines a curated benchmark with a diverse and representative large-scale dataset. Our results show that RefFilter reduces the number of false positives by nearly 32\% on the labeled dataset when compared to a baseline lightweight interference detector. \rb{Mais de um dataset foi usado na pesquisa? Caso afirmativo, melhor a gente ter cuidado para um revisor n\~{a}o achar que o dataset com GT eh ``biased''.}\vl{sim. 2 datasets. ajustei o texto pra citar os 2 datasets e explicar que o benchmark em si é diverso e representativo, evitando se referir apenas ao novo dataset}Although this improvement comes with a non significant increase in false negatives, the overall precision and usefulness of the reports are significantly improved.

Altogether, this paper makes three main contributions: (1) it proposes RefFilter, a refactoring-aware tool for static semantic interference detection that reduces false positives by identifying and discarding refactoring-related alerts; (2) it introduces a novel dataset of 1,087 real-world, diverse merge scenarios to support robust and scalable evaluation;  and (3) it presents an empirical analysis of RefFilter's behavior across both benchmark and large-scale datasets, highlighting the practical impact and limitations of the technique. \rb{Talvez incluir a avalia\c c\~{a}o emp\'{i}rica como uma contribui\c c\~{a}o, real\c cando as implica\c c\~{o}es dos achados aqui tamb\'{e}m.} \vl{boa ideia. adicionei a avalicao empirica como terceira contribuicao.}

%% file: sections/background.tex
\section{Background and Related Work}
\label{sec:background}
Before better motivating the problem we address, we first overview key concepts and related work on semantic conflicts, techniques for detecting them, and refactoring detection tools.
\subsection{Semantic Merge Conflicts}


Although \emph{textual} merge conflicts have been extensively studied, \emph{semantic} conflicts are more challenging. 
They occur when independently correct changes, once combined, lead to unintended behavior deviations~\cite{Pastore2017, shao2009, joana14it}.
As developer intention is hard to rigorously capture, researchers focus on \emph{interference}~\cite{barrosfilho2017using, jesus2024lightweight}, the key concept that lead to the behavior deviations, intended or not. 
Let $B$ be the base version of a program, and let $L$ and $R$ be two sets of independent changes applied to $B$, producing versions $B_L$ and $B_R$. 
Let $M$ denote the merged version that integrates both $L$ and $R$.
Informally, interference occurs when the combined changes in $M$ fail to preserve the behavior established independently by $L$ and $R$, or the unchanged behavior from $B$. 

We formalize interference in terms of state elements modified or observed during program execution.
Let $X$ be the set of all program state elements (e.g., variables, static and instance fields, and arrays). \rb{n\~{a}o entendi data structures aqui \ldots}). \vl{troquei data structures por object fields and arrays para deixar mais claro o que queremos dizer.} For each state element $x \in X$, let $V_B(x)$, $V_L(x)$, $V_R(x)$, and $V_M(x)$ denote the value of $x$ after the execution of versions $B$, $B_L$, $B_R$, and $M$, respectively, under the same initial conditions. \rb{``When executing'' n\~{a}o sugere o momento espec\'{i}fico da observa\c c\~{a}o. Seria como invariantes que devem ser verdadeiras antes ou depois de um m\'{e}trodo espec\'{i}fico ser executado? Em qual momento esse estado eh avaliado? Ap\'{o}s a execu\c c\~{a}o do m\'{e}todo de entry point? Se a vari\'{a}vel for local, como esse estado seria observado? Essa quest\~{a}o s\'{o} est\'{a} surgindo agora porque nos outros artigos eu n\~{a}o lembro de uma especifica\c c\~{a}o mais formal de \emph{interfer\^{e}ncia}, al\'{e}m de indicarmos que vem do artigo de Horwitz et al.. Olhando rapidinho este artigo novamente, ele menciona ``the final value of x \ldots''.} \vl{verdade! a gente ta considerando qualquer momento durante a execução, como ja mencionamos no paragrafo anterior ("modified or observed during execution"). Ajustei o texto pra deixar isso mais explícito, e evitar a impressão de que só estamos olhando o valor final. Valeu pelo toque! \pb{Victor, é só o final mesmo. A notação nem faria sentido se fosse para valores intermediários. Teria que colocar um índice de tempo ou similar. Seria bem mais complicado. Do jeito atual, a limitação é que não funciona com interleaving.}}
We define that changes $L$ and $R$ interfere on state element $x$ if any of the following holds:
\begin{itemize}
    \item \textbf{Type I: Divergent Updates.} All versions produce distinct values for $x$.
    \[
    V_B(x) \neq V_L(x), \quad V_B(x) \neq V_R(x), \quad \text{and} \quad V_L(x) \neq V_R(x).
    \]
    
    \item \textbf{Type II: Non-preserving Integration.} One branch introduces a change to $x$ that is not preserved in the merged version.
   \begin{align*}
    & (V_L(x) \neq V_B(x) \wedge V_L(x) \neq V_M(x)) \\
    & \quad \text{or} \quad (V_R(x) \neq V_B(x) \wedge V_R(x) \neq V_M(x))
\end{align*}

    \item \textbf{Type III: Emergent Divergence.}  Neither branch modifies $x$, but the integration doesn't preserve its original value.
    \[
    V_B(x) = V_L(x) = V_R(x) \quad \text{and} \quad V_M(x) \neq V_B(x).
    \]
\end{itemize}
%
%
\pb{Se precisar de espaço, pode remover de ``We formally...'' até aqui. Isso pode até pegar mal em ICSE, visto como uma formalização excessiva. Já dava para inferir esse último predicado do que tinha sido dito antes. }



\pb{Algumas partes deixei comentadas no .tex. Se sobrar espaço, podemos incluir algumas, mas é provável que precise de ajustes para encaixar bem após as modificações que eu fiz.}

\subsection{Semantic Conflict Detection Techniques}

A number of techniques have been proposed to detect \emph{interference} in software merge scenarios~\cite{silva2017detecting, Da_Silva_2024,barrosfilho2017using, jesus2024lightweight, dejesus2023, Horwitz1989, Sousa2018}. Early research focused on using static analysis based on System Dependence Graphs (SDGs) to capture control and data dependencies across program elements. While SDG-based approaches provide expressive models capable of identifying subtle behavioral interactions, they are computationally expensive and struggle to scale to large codebases.

To address the SDG scalability issues, lightweight static analysis techniques have emerged~\cite{jesus2024lightweight, dejesus2023}. Instead of building full system dependence graphs, these approaches extract localized dependency information directly from the syntax and structure of the code. 
The core idea is to run the analyses in the merged version of the code, which is annotated with metadata indicating instructions modified or added by each developer that contributed to the merge.
The analyses try to explore potential conflicting situations by keeping track of the changes developers make and how they affect state elements. 
This simplification makes semantic interference detection feasible for large-scale and continuous integration scenarios. Lightweight static analysis has demonstrated strong detection capabilities, but remains susceptible to false positives due to its conservative assumptions about potential dependencies. \rb{Talvez citar o trabalho de Susan Horwitz et al. (refer\^{e}ncia [13] atualmente). Tamb\'{e}m incluir que tanto Barros Filho et al. e Horwitz et al. precisam de pelo menos tr\^{e}s vers\~{o}es do software (L, R e B), salvo algum engano. Acho que n\~{a}o precisam de M.Apenas uma sugest\~{a}o, essa se\c c\~{a}o est\'{a} excelente.} \vl{adicionei a referencia de Horwitz. nao entendi bem a parte posterior, ja que a gente nao cita aqui quais versoes cada trabalho usa, seria neste trecho mesmo a sugestão de inclusão das referencias a L, R, B e M?}\pb{Horwitz et al. constroem 4 SDGS, e para o diff, e um do resultado final. Roberto usa só um, o do merge. Mas por mim está bom assim mesmo, sem mencionar esse detalhe.}

A remarkable limitation of the lightweight techniques lies in their inability to distinguish behavior-preserving changes from behavior-modifying changes. In particular, refactorings - code transformations that preserve program semantics - are also annotated as a change that can potentially cause interference, although that only occurs due to behavior changes. This leads to inflated interference reports that burden developers with unnecessary cognitive effort during conflict resolution~\cite{ellis2022operation, oliveira2023do}.

Recent studies highlight the prevalence of refactorings in merge scenarios and their impact on merge effort. Ellis et al.~\cite{ellis2022operation} show that refactorings contribute to larger and more complex \emph{textual} conflicts, while Oliveira et al.~\cite{oliveira2023do} empirically demonstrate that the occurrence of refactorings significantly increases merge effort. These findings underscore the need for interference detection tools that are aware of refactorings to improve precision.


\subsection{Refactoring Detection Tools}


Detecting refactorings accurately is a well-studied problem with applications in software evolution analysis, API migration, regression test selection, and merge conflict resolution~\cite{Tsantalis2022, Tsantalis2018}.
A number of refactoring detection tools have been proposed. 
Among the most prominent are \textit{RefactoringMiner}~\cite{Tsantalis2022, Tsantalis2018} and \textit{RefDiff}~\cite{silva2017detecting}.

\textit{RefactoringMiner} applies fine-grained AST differencing and statement matching algorithms to detect both high-level and submethod-level refactorings. It supports the detection of 40 refactoring types across multiple code element levels and achieves high average precision (99.6\%) and recall (94\%)~\cite{Tsantalis2022, Tsantalis2018}. Its efficiency and accuracy have made it widely adopted in empirical studies and research prototypes.

\textit{RefDiff} combines static analysis with similarity-based heuristics, using adapted TF-IDF weighting to match code entities across versions~\cite{silva2017detecting}. It achieves high precision and strong recall for 13 common refactoring types, offering a scalable \pb{confere? acho que essa é a principal diferenca para o miner.}\vl{suporta Java, JavaScript e C, mas não é independente de linguagem} alternative that performs well across multiple open-source projects.

More recently, \textit{ReExtractor+}~\cite{reextractorplus} introduced advanced entity and statement matching algorithms that leverage reference-based matching to further improve detection accuracy. Its evaluation showed substantial improvements in both false positive reduction and matching granularity compared to previous tools.

While these tools were initially designed for general-purpose refactoring detection, they offer a strong foundation for improving semantic interference detection. By integrating their outputs into interference detection pipelines, it becomes possible to identify which reported interferences are spurious, artifacts of behavior-preserving refactorings. In a preliminary evaluation, we assessed different refactoring detection tools and selected RefactoringMiner and ReExtractor+ based on their precision, stability, and support for a broad range of refactoring types in Java projects. Our work leverages both RefactoringMiner and ReExtractor+ to build a refactoring-aware interference detection tool capable of filtering out such cases, thereby significantly improving precision without substantial loss in recall.

\rb{Talvez s\'{o} justificar, brevemente, o porqu\^{e} de ter escolhido RefactoringMiner e ReExtractor+. Talvez incluir que, durante uma avalia\c c\~{a}o preliminar, essas ferramentas se mostraram mais adequadas. Descrever minimamente tal avalia\c c\~{a}o preliminar.}
\vl{Boa. coloquei uma justificativa curta, mencionando que fizemos uma avaliação preliminar e escolhemos com base em precisão, estabilidade e cobertura. Isso já dá um respaldo melhor no texto.}

%% file: sections/problem-definition.tex
\section{Motivating Example and Problem Definition}
\label{problem-definition}

\begin{figure*}[ht]
    \centering
    \includegraphics[width=0.90\textwidth]{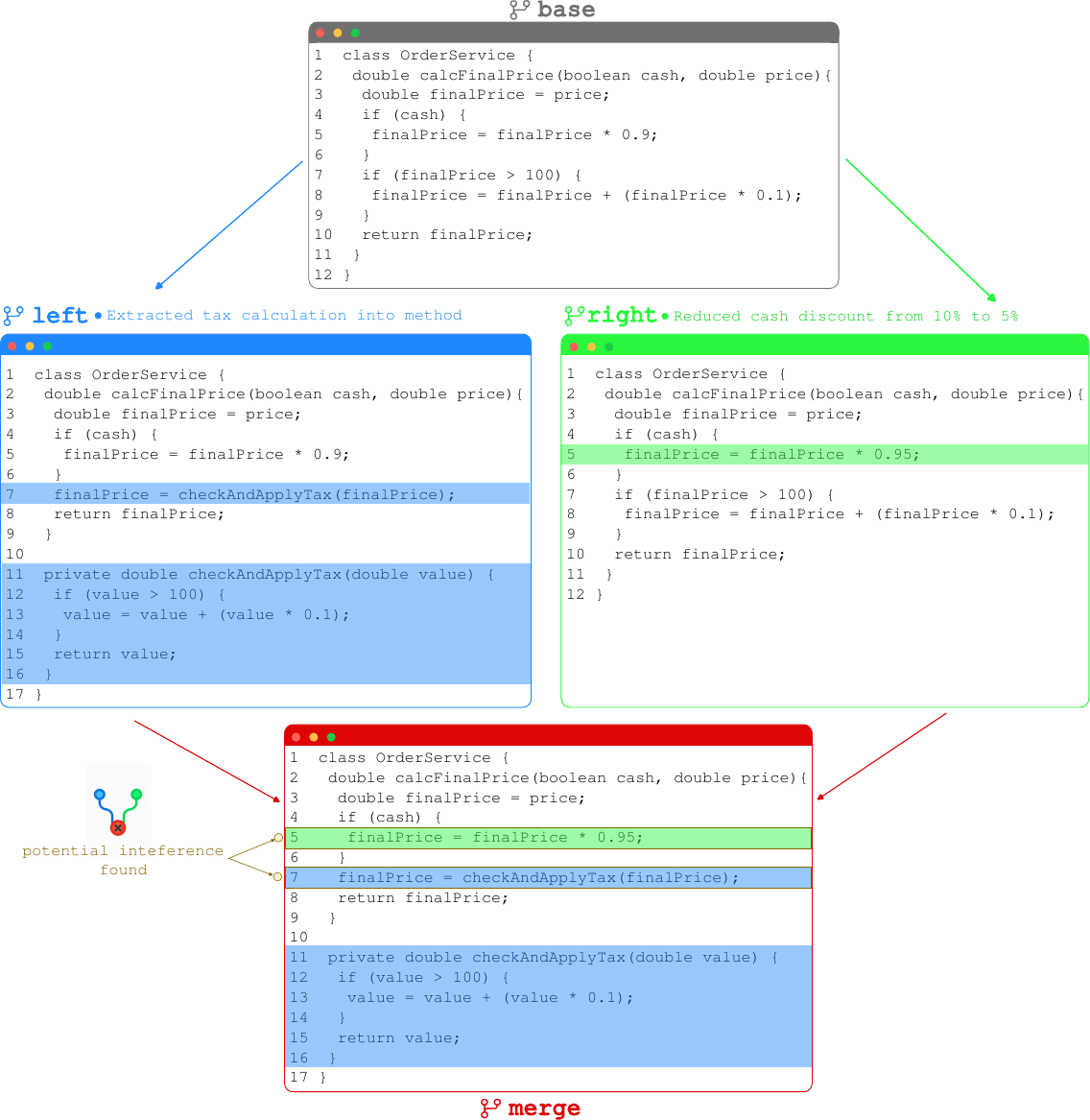}
    \caption{Merge scenario with refactoring (left) and business rule change (right). \pb{Tá bem pequena a figura. Tem como aumentar? o Fonte do código aqui deveria ser do mesmo tamanho do fonte do texto.} \pb{Remover a linha de potential interference found para a linha 13 no codigo de merge.\vl{feito. aumentei a figura e removi a referência para a linha 13}}}
    \label{fig:problem-example}
\end{figure*}

To illustrate how static analysis interference detection tools might report false positives, consider the method \emph{calculateFinalPrice} in class \emph{OrderService}, as shown in Figure~\ref{fig:problem-example}. 
In this example, two branches independently modify different parts of the method. 
The \emph{left} branch applies a behavior-preserving refactoring by extracting tax calculation logic into a separate method. 
On the other hand, the \emph{right} branch modifies the business rule by changing the discount calculation policy. 
Since the edits affect different regions of the method, no textual conflict is reported during the merge. 

However, a lightweight static semantic interference detection tool~\cite{jesus2024lightweight,dejesus2023} analyzes data dependencies and reports a potential interference in this merge: 
%
\[
PI_1 = \{(\texttt{OrderService}, 5), (\texttt{OrderService}, 7)\},
\]\pb{Tirei $(\texttt{OrderService}, 13)$ do conjunto acima. Confere com o log gerado pela ferramenta. Pode ter sido alguma confusão tua ou um bug da ferramenta, mas não vejo como faria sentido ter isso no conjunto.} \vl{ok}
%
%
indicating that the change in line 5 in the merge potentially interferes with the change in line 7. 
The tool conservatively reports this because it detects a data flow from 5 to 7, which would be a problem if left changes were assuming for \texttt{finalPrice} the calculation logic in the base version, whereas the merge contains the new logic coming from right. 
In this case, however, there is no interference since left simply refactored the code, which makes no assumptions about the logic in other parts of the code.
A refactoring-aware interference detection tool would not report this potential interference, as it recognizes that the change in line 7 is behavior-preserving, saving developer effort.


To precisely formulate this problem, we consider that interference detection tools report a set of \emph{potential interferences},
denoted by $\mathcal{PI} = \{ PI_1, PI_2, \dots, PI_n \}$, where each $PI_j$ represents a set of
pairs $(C, l)$, with $C$ denoting a class and $l$ a line number (in the merge version\pb{Confere?}\vl{sim}) involved in the reported interference $j$. \rb{N\~{a}o estou certo que eu entendi essa formula\c c\~{a}o. Seria legal discutirmos rapidinho}\pb{Tá confuso isso aqui, Victor. Não seria simplesmente $(C_j, l_{j})$. Não vi porque precisa mais do que isso.}\vl{simplifiquei mais. como não era necessario realmente poluir a notação, tirei o $i$ e $j$ do par classe/linha }
%
%
%
%
Let now 
\begin{itemize}
    \item $L \subseteq \mathcal{C} \times \mathbb{N}$ be the set of pairs $(C, l)$ such that class $C$ and line $l$ were modified by the \emph{left} commit; \pb{o $c$ em $L$ parece ser desnecessário, complica a notação. Por que não só $L$? A mesma coisa vale para $R$ abaixo. Além de complicar, usar $c$ aqui dificulta mais abaixo tb, quando poderia chamar commit de $c$, mas chama de $f$.} \vl{ o $c$ é pra indicar changes. Left changes e Right changes. mas vou simplificar.}
    \item $R \subseteq \mathcal{C} \times \mathbb{N}$ be the set of pairs $(C, l)$ such that class $C$ and line $l$ were modified by the \emph{right} commit;
    \item $R_f(C, l)$: a predicate that holds if the modification at $(C, l)$ corresponds to a behavior-preserving refactoring. \pb{o $f$ em $R_f$ parece ser desnecessário. $f$ de que?}
\end{itemize}
Considering our example, $R$ would be $\{(\texttt{OrderService}, 5)\}$ and $L$ would contain seven pairs, one with line 7 and the others with lines 11-16, all line numbers referring to the merge version of the code. \pb{Precisando de espaço, remove essa parte que ilustra os conceitos. Não é muito necessária.}
%

Finally, we define the predicate $\Psi(c, PI_j)$ that specifies whether all modifications made by a given commit $c \in \{L, R\}$ \pb{Ruim chamar um commit de $f$, ao invés de $c$. Ruim tb porque $f$ é usado em $R_f$ mas acima. Gera confusão.} involved in a potential interference $PI_j$ are refactorings:
%
%
\begin{equation}
\Psi(c, PI_j) \equiv \forall (C, l) \in PI_j:\; 
\Big[\, (C, l) \notin c \nonumber \vee\; R_f(C, l) \,\Big].
\label{eq:refactoring-predicate} 
\end{equation}
%
In other words, $\Psi(c, PI_j)$ holds if either (i) the pair $(C, l)$ was not modified by $c$, or (ii) it was modified and the change is classified as a refactoring. We now formally define a \textbf{refactoring-induced false positive}.
\begin{definition}[False Positive of Interference due to Refactoring]
\label{def:false-positive}
A potential interference $PI_j$ is classified as a false positive caused by refactoring if
\begin{equation}
\Psi(L, PI_j) \vee \Psi(R, PI_j).
\label{eq:false-positive-condition}
\end{equation}
\end{definition}
If for at least one branch all modifications related to $PI_j$ correspond to refactorings (or are unrelated), then $PI_j$ is considered spurious. Therefore, the problem addressed in this work is the following: given a set of potential interferences $\mathcal{PI}$ reported by a static interference detection tool, identify and discard all $PI_j \in \mathcal{PI}$ that satisfy Equation~\eqref{eq:false-positive-condition}.
By doing so, we aim to reduce the number of false positives in semantic interference reports while preserving actual interferences relevant to developers. 
%
%

%

Now we to apply the $\Psi$ predicate (Equation~\ref{eq:refactoring-predicate})\pb{A referência para a equação não parece estar funcionando aqui}\vl{ajustei} to evaluate whether $PI_1$ corresponds to a refactoring-induced false positive in our motivating example. 
As $R_f(\texttt{OrderService}, 7)$ holds, and $\Psi(L, PI_1)$ also holds, we confirm a false positive according to Definition~\ref{def:false-positive} (Equation~\eqref{eq:false-positive-condition}).



%% file: sections/proposed-method.tex
\section{RefFilter}

This section presents the refactoring-aware semantic interference detection technique  implemented by RefFilter. While building upon a prior static interference detection technique, we here introduce a novel and modular filtering stage based on refactoring awareness, which enables a significant reduction of false positives caused by behavior-preserving refactorings. This integration results in a more precise and practical semantic conflict detection technique. \rb{Esse par\'{a}grafo pode levar um revisor a achar que a contribui\c c\~{a}o \'{e} incremental. Eu n\~{a}o acho que isso prejudica o trabalho, mas ICSE tem se tornado uma confer\^{e}ncia super complicada.} \vl{verdade! Ajustei pra reforçar que, apesar de usar uma tecnica existente como base, a nossa contribuição é original e relevante - com um estágio novo e modular de filtragem que melhora significativamente a precisão. Isso ajuda a evitar que o revisor leia como algo meramente incremental}\pb{Acho que desta vez podemos tentar desta forma porque tá muito em cima, mas a narrativa melhor seria vender o pipeline todo, aqui dá muito a entender que a contribuição é só a filtragem.}

\subsection{Overview of the Technique}
\label{sec:overview}

Given a target merge commit, the RefFilter technique proceeds in two main phases: (i) static interference detection, and (ii) refactoring-aware filtering.

\subsubsection*{Phase 1: Static Interference Detection.} 
Initially, an existing static semantic interference detection tool is applied to the merged version of the code. This tool analyzes the integrated changes from both the \emph{left} and \emph{right} branches and identifies a set of potential interferences $\mathcal{PI} = \{ PI_1, PI_2, \dots, PI_n \}$, as defined in
Section~\ref{problem-definition}. 

If no potential interferences are reported (i.e., $\mathcal{PI} = \emptyset$), the analysis terminates, and the merge is considered interference-free. Otherwise, the reported potential interferences are passed to the refactoring-aware filtering phase.

\subsubsection*{Phase 2: Refactoring-Aware Filtering.}
For each potential interference $PI_j \in \mathcal{PI}$ reported by the static analysis tool, \emph{RefFilter} evaluates whether the interference is a legitimate case or a false positive caused by refactorings.

To perform such an evaluation, automated refactoring detection tools (such as \emph{RefactoringMiner} and \emph{ReExtractor+}) are applied to both branches to classify which modifications correspond to refactorings. 
These tools, however, return refactoring information in terms of line numbers in the parents commits, not in the merge commit. 
As the interference information from the first phase is based on line numbers from the merged version of the code, we apply a line alignment algorithm. \pb{Tentei vender melhor o peixe aqui pois realmente tem uma dificuldade extra. Com isso, pode parecer menos incremental. Vê se faz sentido e se tem como vender melhor.}

\begin{figure}[ht]
    \centering
    \includegraphics[width=0.95\linewidth]{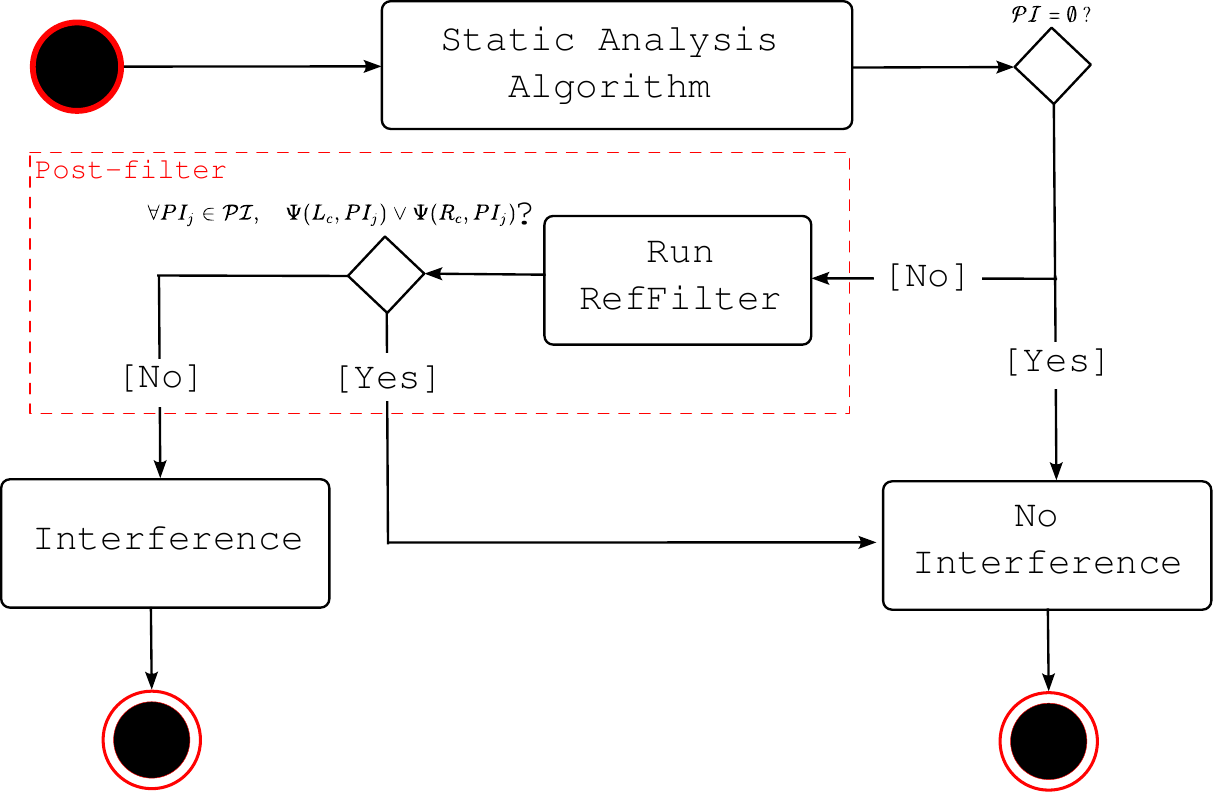}
    \caption{Overview of \emph{RefFilter} workflow.}
    \label{fig:method}
\end{figure}

Based on this information, \emph{RefFilter} applies the predicate $\Psi(c, PI_j)$ defined in Equation~\ref{eq:refactoring-predicate} to assess whether, for each reported interference, all the changes made by either $L$ or $R$ correspond exclusively to refactorings. If this condition holds for either side (i.e., if Equation~\ref{eq:false-positive-condition} evaluates to true), the potential interference $PI_j$ is classified as a false positive due to refactoring and discarded. Otherwise, $PI_j$ is finally reported as a true potential interference to the user. The complete workflow of the approach is illustrated in Figure~\ref{fig:method}. \pb{verificar os números das equações aqui, acho que estão com problema.}

\subsection{Refactoring-Aware Semantic Conflict Detection Algorithm}
\label{sec:algorithm}

This section presents the complete algorithm implemented by \emph{RefFilter} to classify potential semantic interference as a false positives caused by refactorings.
The algorithm receives as input the repository URL, the merge commit $M$, and the set of potential interferences $\mathcal{PI}$ reported by the static interference detection tool. As output, the algorithm returns \textbf{False} if all potential interferences are classified as false positives caused by refactorings; otherwise, it yields \textbf{True}. The detection process consists of three stages: extraction of modified lines, refactoring detection, and interference filtering.

\todo[inline]{Revisar o pr\'{o}ximo par\'{a}grafo, pois ajustei conforme o meu entendimento (rb)}
\vl{revisei, o par\'{a}grafo ficou claro e objetivo.}

In the first stage, the algorithm extracts the sets of modified locations, $L$ and $R$, from the merge commit $M$. Each set contains all class-line pairs $(C, l)$ modified by the left and right branches, respectively.
Note that a merge commit may include multiple intermediate commits between the base and each branch head. To simplify diff computation and align file versions for subsequent analysis, we squash the changes from each branch into a single virtual commit.
After squashing, line numbers may still differ between $M$ and its parents. Therefore, for each pair $(C, l) \in L \cup R$, a content-based mapping is performed to locate the corresponding line numbers in the squashed left and right branches. This mapping employs approximate string matching based on the Jaro-Winkler similarity metric~\cite{Winkler1990}, selecting candidates according to highest similarity and minimal positional distance.

Following the mapping, the algorithm independently performs refactoring detection for $L$ and $R$. The function \emph{{DetectRefactorings}(C)}, detailed in Algorithm~\ref{alg:refdetect}, executes two state-of-the-art refactoring detection tools. The first tool, \emph{RefactoringMiner}~\cite{Tsantalis2022, Tsantalis2018}, employs AST differencing and supports 40 refactoring types with high precision and recall. During the development of this work, RefactoringMiner introduced the \texttt{PurityChecker} API, which improves the detection of behavior-preserving (pure) refactorings. Given the central role of distinguishing such refactorings in our filtering strategy, we re-executed all experiments using the updated version. The second tool, \emph{ReExtractor+}~\cite{reextractorplus}, uses reference-based entity matching, improving detection of nested and cross-cutting refactorings. The union of the results produced by both tools yields comprehensive refactoring sets $RefL$ and $RefR$ for the left and right branches, respectively.

In the final stage, for each potential interference $PI_j \in \mathcal{PI}$, the algorithm applies the refactoring-aware filtering predicate $\Psi$ defined in Equation~\ref{eq:refactoring-predicate}. Specifically, the classification rule from Definition~\ref{def:false-positive} is evaluated by computing $\Psi(L, PI_j) \vee \Psi(R, PI_j)$. \pb{Aqui $\Psi$ tem 3 parâmetros, lá na seção anterior tinha 2. Verificar isso e o número da equação.}\vl{é pq na implementação eu passei o parametro adicional ao invés de computar o valor de $RefL$ no proprio $\Psi$. de todo jeito, ajustei pra ficar igual ao que está na formulação, pra nao gerar essa duvida} If this expression evaluates to true for any branch, the interference $PI_j$ is classified as a false positive due to refactorings. \pb{Tirei desta seção todo tipo de texto na linha de ``otherwise, it is classified as a real interference''. Na verdade, RefFilter corta FP. Mas se ele não cortar, pode ainda ser FP por outro motivo. Não há garantia que seja TP! Verificar se não há esse tipo de texto no resto do artigo} The entire procedure is formally described in Algorithms~\ref{alg:refilter}--\ref{alg:refdetect}. \pb{Não revisei os algoritmos, assumi que tinhas feito isso com cuidado.}

\begin{algorithm}[ht]
\caption{RefFilter Interference Classification Algorithm}
\label{alg:refilter}
\begin{algorithmic}[1]
\Require Repository URL, merge commit $M$, potential interferences $\mathcal{PI}$
\Ensure \textbf{True} if real interference exists, \textbf{False} otherwise

\State $L, R \gets \mathsf{ExtractModifiedLines}(M)$
\State $B, L, R \gets \mathsf{GetParents}(M)$
\State $\mathsf{Squash}(Base \rightarrow Left)$, $\mathsf{Squash}(Base \rightarrow Right)$
\State $L \gets \mathsf{MapLines}(L, Left)$
\State $R \gets \mathsf{MapLines}(R, Right)$

\ForAll{$PI_j \in \mathcal{PI}$}
    \State $pass_L \gets \Psi(L, PI_j)$
    \State $pass_R \gets \Psi(R, PI_j)$
    \If{\textbf{not} ($pass_L \vee pass_R$)}
        \State \Return \textbf{True}
    \EndIf
\EndFor

\State \Return \textbf{False}
\end{algorithmic}
\end{algorithm}

\begin{algorithm}[ht]
\caption{Predicate $\Psi(F_c, PI_j)$}
\label{alg:psi}
\begin{algorithmic}[1]
\Require Modification set $F_c$, interference $PI_j$
\Ensure \textbf{True} if predicate holds
\State $RefF \gets \mathsf{DetectRefactorings}(F_c)$
\ForAll{$(C, l) \in PI_j$}
    \If{$(C, l) \in F_c \wedge (C, l) \notin RefF$}
        \State \Return \textbf{False}
    \EndIf
\EndFor
\State \Return \textbf{True}
\end{algorithmic}
\end{algorithm}

\begin{algorithm}[ht]
\caption{DetectRefactorings($C$)}
\label{alg:refdetect}
\begin{algorithmic}[1]
\Require Commit hash $C$
\Ensure Set of refactorings $RefSet$

\State $RefSet \gets \emptyset$
\ForAll{$T \in \{\mathsf{RefactoringMiner}, \mathsf{ReExtractor+}\}$}
    \State $R \gets T(C)$
    \State $RefSet \gets RefSet \cup R$
\EndFor
\State \Return $RefSet$
\end{algorithmic}
\end{algorithm}

%% file: sections/experimental-setup.tex
\section{Experimental Setup}
\label{sec:experimentalsetup}

\label{sec:researchquestions}

To evaluate whether RefFilter effectively reduces false positives in interference detection while preserving recall, we focus on three research questions that directly reflect our main goals: reducing false positives caused by refactorings, and understanding the trade-offs in terms of recall. Specifically, we address the following questions:
\begin{itemize}
    \item \textbf{RQ1:} To what extent does RefFilter reduce false positives compared to traditional static analysis techniques?
    \item \textbf{RQ2:} To what extent does the reduction in false positives affect the number of false negatives?
    \item \textbf{RQ3:} To what extent do the improvements achieved by RefFilter generalize to large-scale, diverse merge scenarios?
\end{itemize}
\pb{A gente não tem nada de análise de tempo, para saber se o custo computacional da técnica é aceitável? Se tiver, mesmo que básico seria legal colocar aqui algo como: We also evaluate RefFilter's computational cost in a preliminar way, just to make sure that the technique is not computationally prohibitive.}\vl{coloquei}

We also conduct a preliminary evaluation of RefFilter’s computational cost to ensure it is not prohibitive.
RQ1 and RQ2 are evaluated using two complementary datasets. The first is a benchmark dataset with 99 merge scenarios and ground truth, which allows precise measurement of true and false positives and negatives. The second is a large-scale dataset with 1,087 merge scenarios from diverse real-world projects, used to assess whether the trends observed in the benchmark persist at scale (RQ3). Together, these datasets provide a balance between evaluation depth and breadth.

\todo[inline]{Talvez incluir uma justificativa para essas quest\~{o}es de pesquisa. Os dois datasets s\~{a}o usados para responder a ambas quest\~{o}es de pesquisa? N\~{a}o seria legal incluir uma quest\~{a}o relacionada \`{a} acur\'{a}cia das ferramentas de detec\c c\~{a}o de refactoring?}
\vl{Ajustei o texto pra incluir a justificativa das RQs e esclarecer que usamos os dois datasets de forma complementar: o com GT pra medir com precisão os acertos e erros, e o grande pra ver se os resultados se mantêm em larga escala. Sobre a sugestão de incluir uma RQ sobre acurácia, optei por não colocar porque a técnica não busca aumentar os TPs, mas sim reduzir os FPs e analisar o impacto disso — por isso focamos nas duas perguntas que realmente refletem nossos objetivos principais.}

\subsection{Datasets}
\label{sec:datasets}

To evaluate our technique, we designed two complementary experiments based on the two distinct datasets we mentioned before and detail now. \rb{Caso os datasets explorem quest\~{o}es de pesquisa distintas, deixar isso claro neste ponto.} \vl{Adicionei um paragrafo no item 5.1 Research Questions (both questions are...) para explicar melhor a relacao dos datasets com as RQs.}

\noindent\subsubsection*{Experiment 1: Performance Comparison with Existing Methods}
For Experiment 1, we employed the same benchmark dataset previously used in related work~\cite{jesus2024lightweight, dejesus2023}, which consists of 99 merge scenarios. For each scenario, a ground truth label is available indicating whether interference actually exists. This allows for direct performance comparison between RefFilter, a pure static analysis interference detection tool, and random baseline classifiers using standard evaluation metrics.
\todo[inline]{Tendo espa\c co, talvez seja legal incluir mais informa\c c\~{o}es sobre este datasets\ldots. Consultar o paper do Matheus.}
\vl{é... o espaco ta na conta mesmo. talvez ate esteja passando um pouco apos esses ajustes}

\noindent\subsubsection*{Experiment 2: Scaling and Diversity Evaluation}

\begin{figure*}[ht]
    \centering
    \includegraphics[width=0.65\textwidth]{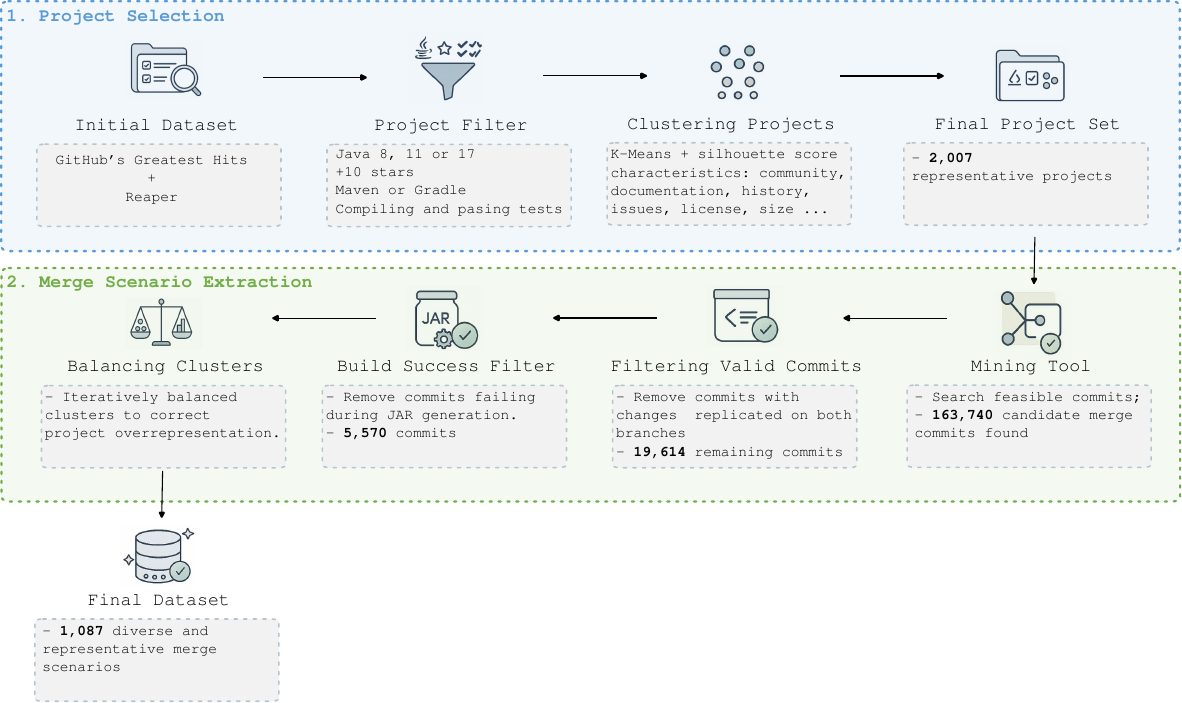}
    \caption{Dataset construction workflow.}
    \label{fig:dataset}
\end{figure*}

\todo[inline]{Esse dataset foi muito bem projetado. Talvez fosse interessante public\'{a}lo como dataset em alguma trilha de MSR.}
\vl{otima ideia! Podemos fazer isso sim.}
To evaluate RefFilter's scalability and performance across diverse projects and merge scenarios, we constructed a novel dataset comprising 1,087 merge commits. The dataset creation involved two main stages: (i) project selection and (ii) merge scenario extraction. Figure~\ref{fig:dataset} shows the dataset construction workflow.

In the project selection stage, we followed a strategy similar to~\cite{Schesch}, initially leveraging Reaper~\cite{munaiah2017reaper} and GitHub's Greatest Hits~\cite{githubGreatestHits} datasets, which contain high-quality open-source repositories. From these, we selected Java projects with at least 10 stars \rb{parece pouco, existem projetos em [22] com menos de 10 estrelas?}\vl{nao verifiquei explicitamente quantos repositorios passam individualmente me cada um dos filtros. esses filtros foram adotados com esses parametros por conta do trabalho de Schesch, que conseguiu reunir uma sublista de projetos java que geravam o jar automaticamente } that used Maven or Gradle for build automation. Only projects where the latest commit on the main branch compiled and passed tests within 30 minutes (using JDK 8, 11, or 17) were included, ensuring feasibility for automated \texttt{.jar} artifacts generation, which are necessary for running static analyses. The projects were clustered using the K-Means algorithm~\cite{MacQueen1967}, considering multiple project-level features (architecture, community, documentation, development history, issue tracking, license, size, unit test presence, and GitHub popularity via stars) in order to ensure representativeness across heterogeneous profiles. The optimal number of clusters was selected based on the silhouette coefficient~\cite{Silhouettes}, a density-and separation-based internal validation metric that estimates cluster quality by comparing intra-cluster cohesion and inter-cluster separation. After filtering, 2,007 projects remained in the final selection.

In the second stage, we employed the Mining tool~\cite{mining} to extract merge commits where both parents modified the same method within the same Java class.
These are easier to manually evaluate, in case interference is reported.
Running interference detection static analysis tools in other kinds of scenarios is also more expensive, as we have to compute appropriate entrypoints. 
\rb{verificar a justificativa desta restri\c c\~ao antes no paper, e refor\c car aqui} \vl{ok. nao achei nos papers anteriores a justificativa tecnica disso, entao ajustei apenas para mencionar que é uma limitacao das ferramentas de analise estatica base adotadas}. This yielded 163,740 candidate merge commits. We observed that in many cases, changes on both parents occurred on the same lines--- likely resulting from common ancestor changes replicated on both branches. Such cases were discarded, reducing the dataset to 19,614 commits. Commits failing during automated release generation (including jar builds) were further excluded, leaving 5,570 commits. Finally, to correct for project overrepresentation, we iteratively balanced cluster representation, resulting in the final dataset containing 1,087 diverse and representative merge scenarios.

\subsection{Evaluation Methodology}
\label{sec:methodology}


Our empirical assessment consists of two experiments. \textbf{In the first (Experiment 1)}, both pure static analysis and RefFilter were applied to the 99
benchmark scenarios. Performance was evaluated using standard confusion matrix metrics: precision, recall, accuracy, and F1-score.

Since both techniques operate on the same merge scenarios, observations are paired. To assess whether differences in false positives (RQ1) and false negatives (RQ2) were statistically significant, we applied McNemar's exact test~\cite{McNemar1947}.

In order to establish lower-bound performance references, we also simulated two random classifiers: (i) a \textit{coin-flip classifier} that randomly assigns interference labels with 50\% probability, and (ii) a \textit{calibrated random classifier} that assigns positive labels with probability matching the empirical prevalence of interference in the dataset (28\%, based on the proportion of true interference cases observed in the benchmark dataset). Comparisons with these baselines were performed using empirical non-parametric hypothesis testing, applying Monte Carlo simulation~\cite{efron1994} with 10,000 iterations and computing p-values based on the proportion of random simulations exceeding the observed classifier performance. \rb{Achei super interessante, apenas n\~{a}o entendi a escolha de 28\%} \vl{A escolha dos 28\% foi baseada na proporção real de interferências verdadeiras observadas no dataset com ground truth. Adicionei essa justificativa direto no texto pra esclarecer.}

Beyond quantitative evaluation, we manually analyzed cases where RefFilter correctly or incorrectly classified merge scenarios to better understand patterns, strengths, and limitations of the approach.


\todo[inline]{Acho que dever\'{i}amos discutir as quest\~{o}es de pesquisa. N\~{a}o est\'{a} claro para mim como o experimento 2 ajuda a respond\^{e}las.}
\vl{concordo. adicionei um parágrafo explicando melhor o papel do Experimento 2. Ele serve justamente pra reforçar que as reduções de falsos positivos observadas no primeiro experimento não são casuísticas, mas sim consistentes mesmo em um dataset maior, diversificado e representativo do mundo real. Isso dá mais robustez pras respostas das RQs 1 e 2.o Parágrafo é o "This second experiment complements..."}

Due to the absence of interference ground truth in the larger dataset, in the second experiment (Experiment 2) we focused our evaluation on RefFilter's ability to correctly discard false positives caused solely by refactoring modifications. All merge scenarios discarded by RefFilter as refactorings were manually reviewed to verify whether they corresponded to true refactorings. 
Each scenario was reviewed by two reviewers using pair-reviewing.

This second experiment complements the first by assessing whether the results observed in the benchmark dataset generalize to a broader and more diverse set of merge scenarios. While Experiment 1 precisely quantifies false positives and false negatives using ground truth, Experiment 2 verifies if similar reductions in false positives hold across large-scale real-world scenarios. This setup strengthens the validity of our answers to RQ1 and RQ2, ensuring that the observed gains are not limited to a small or potentially specific dataset.

Importantly, we did not manually validate scenarios that RefFilter classified as interferences, nor did we determine whether non discarded refactorings contained actual interferences. Thus, this evaluation assesses RefFilter’s precision in refactoring detection but does not provide recall estimates for the overall dataset, as this would be extremely hard due to the dataset size and the nature of the manual analysis process, which requires deep semantic understanding of the code.

Similar to Experiment 1, we also manually inspected selected cases of correct and incorrect decisions by RefFilter to extract insights regarding recurring patterns and challenging situations.

\todo[inline]{Acredito que RefDiff seja independente da ferramenta de detec\c c\~{a}o de interfer\^{e}ncia. De qualquer forma, senti falta aqui em settings de sermos mais expl\'{i}citos sobre como executamos / configuramos as ferramentas de an\'{a}lise est\'{a}tica. Talvez indicar que executamos nos mesmos settings de trabalhos anteriores, e resumir os principais itens de configura\c c\~{a}o. Seria o caso de pedir para Galileu o Matheus escreverem algo sobre isso?}
\vl{entendi. vou esperar entao a opiniao do prof. paulo pra ver se sugere pedir essa inclusao pra matheus ou galileu}\pb{Acho que pode ficar para a próxima mesmo, mas realmente faltou um pouco mais de detalhe sobre a SA, a ferramenta base.}

%% file: sections/experimental-results.tex
\section{Experimental Results}
\label{sec:results}

\todo[inline]{Melhor revisar as partes anteriores do paper para indicar que usamos essa vers\~{a}o com PurityChecker. Eu removeria esse par\'{a}grafo introdut\'{o}rio daqui.}
\vl{Ok. Removi aqui e adicionei essa questão na seção que apresento o algoritmo "DetectRefactorings". Coloquei "During the development of this work, RefactoringMiner introduced the PurityChecker API, which improves the detection of behavior-preserving (pure) refactorings. Given the central role of distinguishing such refactorings in our filtering strategy, we re-executed all experiments using the updated version"}

We now discuss the results of our experiments.
We first present the results of Experiments~1 and 2, followed by findings from our qualitative analysis of RefFilter's filterings.

\subsection{Performance Comparison with Existing Methods}
\label{sec:exp1results}

Table~\ref{tab:confusion} presents the confusion matrices obtained for each evaluated approach, comparing RefFilter with the baseline of a pure static analysis (SA) \rb{acho interessante refor\c car o dataset usado aqui\ldots Pode ser iniciando a pr\'{o}xima senten\c ca como: Recall that the ABC dataset, used to investigate RQ1, includes precise \ldots} \vl{Adicionei uma menção direta ao dataset de 99 cenários com ground truth, como você indicou, pra reforçar melhor o contexto da tabela.}. As mentioned earlier, this first experiment uses the benchmark dataset with 99 merge scenarios and ground truth annotations.
\begin{table}[h]
\centering
\caption{Confusion Matrices (Experiment 1)}
\label{tab:confusion}
\begin{tabular}{lcccc}
\hline
\textbf{Classifier} & \textbf{TP} & \textbf{FP} & \textbf{FN} & \textbf{TN} \\
\hline
SA & 15 & 32 & 13 & 39 \\
SA + RefFilter & 14 & 22 & 14 & 49 \\
\hline
\end{tabular}
\end{table}

RefFilter significantly reduces false positives, from 32 (in pure static analysis) to 22, while maintaining a similar number of true positives (15 vs. 14) and false negatives (13 vs. 14). This reduction in false positives--- representing a relative decrease of 31.2\%--- provides evidence to answer RQ1. This reduction is especially relevant in the context of software merging, where each false positive implies unnecessary manual inspection. By filtering out spurious interference caused by refactoring changes, RefFilter alleviates developer burden and improves the utility of static analysis techniques for interference detection.

Table~\ref{tab:metrics}  shows the performance metrics, and the analysis reveals that the RefFilter technique demonstrates clear improvements, particularly in precision and overall classification balance, compared to pure static analysis (SA). Specifically, precision increased from 0.319 to 0.389, directly reducing the developer effort spent on false positives, a core motivation of this work. Although the recall decreased slightly from 0.536 to 0.500, this variation was not statistically significant (McNemar p = 1.0), indicating that the gain in precision was achieved without compromising the sensitivity of the technique. As a consequence, the F1-score improved from 0.400 to 0.438, reflecting a better balance between precision and recall.

\begin{table}[h]
\centering
\caption{Performance Metrics (Experiment 1)}
\label{tab:metrics}
\begin{tabular}{lcccc}
\hline
\textbf{Classifier} & \textbf{Precision} & \textbf{Recall} & \textbf{Accuracy} & \textbf{F1-score} \\
\hline
SA & 0.319 & 0.536 & 0.545 & 0.400 \\
SA + RefFilter & 0.389 & 0.500 & 0.636 & 0.438 \\
\hline
\end{tabular}
\end{table}
\pb{Na dúvida se a gente deveria apresentar a acurácia aqui. A gente pode estar levantando a bola para um revisor cortar, já que os números não são tão bons, mesmo com RefFilter. A RQ1 já foi respondida com a redução de falsos positivos. A gente parece mais perder que ganhar com isso aqui. Opção 1: deixar como está e arriscar. Opção 2: remover só a tabela com as métricas de acurácia, deixar o texto. Opção 3: remover tudo, tabela e texto, e focar apenas em reducao de FPs, nao o impacto disso em acurácia, ou só dizer que a redução em FP deu um aumento de X\% em precision, Y\% em recall, etc. sem mostrar as medidas originais.}
\vl{como o professor ficou na dúvida, vou manter a seção, já que desde começo da pesquisa tínhamo interesse em coletar essa métrica, acho que faz sentido apresentar.}

To rigorously assess whether these improvements could be attributed to chance, we applied statistical hypothesis tests. McNemar's exact test confirmed that RefFilter achieved a statistically significant reduction in false positives compared to static analysis (\textit{p} = 0.00195), providing strong evidence that RefFilter addresses the primary challenge targeted by this work (RQ1). Additionally, no statistically significant difference was found regarding false negatives (\textit{p} = 1.0), supporting RQ2. \rb{Talvez fosse interessante incluir mais quest\~{o}es de pesquisa, pois estamos respondendo tanto RQ1 quanto RQ2 no experimento 1. J\'{a} tinha deixado uma observa\c c\~{a}o nesta linha antes.} \vl{tava pensando sobre isso. pode ser mesmo, mas acho que não precisamos criar mais RQs por conta do outro experimento. ele serve justamente pra validar se os resultados das RQs 1 e 2 se mantêm fora do dataset com GT. Mantemos o foco nas mesmas RQs, mas com evidência mais robusta}

\todo[inline]{Seria o caso de incluir uma RQ relacionada \`{a} tempo de execu\c c\~{a}o? Esse tempo inclui tanto o tempo de static analysis quanto o tempo de identifica\c c\~{a}o de refactoring? Talvez fosse importante deixar isso claro.}
\vl{boa pergunta... esclareci que o tempo se refere especificamente à execução do RefFilter, e não da análise estática completa. em relacao à sugestão de incluir uma RQ sobre desempenho, seria legal futuramente — mas como não planejamos os experimentos com essa pergunta em mente (e não capturamos os tempos separados), precisaríamos reexecutar tudo com ajustes no código. Então por ora acho que seria o caso de submeter sem incluir essa nova RQ}

\begin{figure}[ht]
    \centering
    \includegraphics[width=0.85\linewidth]{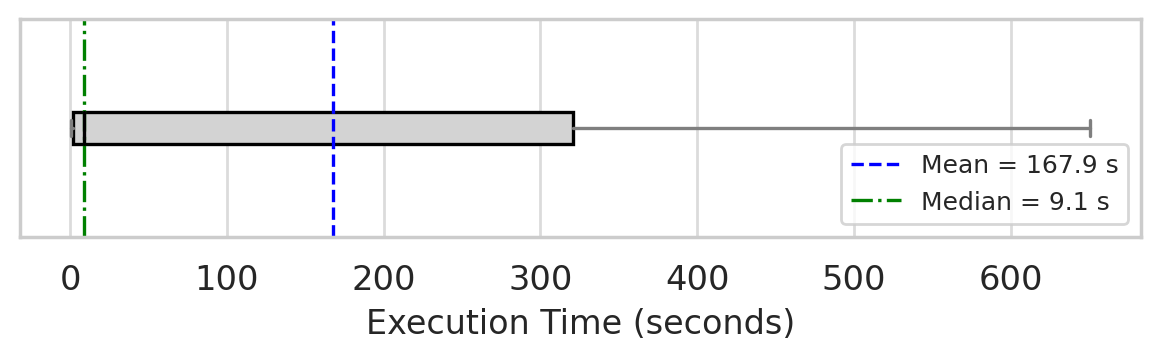}
    \caption{Execution time per scenario.}
    \label{fig:times}
\end{figure}

Figure~\ref{fig:times} presents a detailed boxplot of the execution time (in seconds) across all experimental scenarios, measured specifically for the execution of the RefFilter filtering phase. All experiments were executed on a machine running Ubuntu 20.04.6 LTS (64-bit), with 16GB of RAM and a 12-core Intel® Core™ i7-1255U processor. The distribution is right-skewed, with a median of 9.1 seconds and a wide interquartile range (1.7–321 s), reflecting expected variability. This variation arises from the intrinsic differences in scenario complexity--- ranging from small projects or simple merges to large projects and commits with many modified files which have to be analyzed for refactoring.\pb{Para o futuro, não sei se a gente podia rodar a ferramenta de refactoring só após rodar a análise estática, apenas nos arquivos reportados pela análise} While the mean reaches 168 seconds, it is affected by a few high-duration cases, as confirmed by the non-normality observed in the Shapiro-Wilk test~\cite{shapiro} (p < 0.001).

\todo[inline]{Acho que essa discuss\~{a}o do pr\'{o}ximo par\'{a}grafo deveria vir antes da discuss\~{a}o sobre tempo de execu\c c\~{a}o. Come\c caria com:
One might argue that the performance of SA + RefFilter remains unsatisfactory and is statistically comparable to random guessing. However, \ldots}

To further assess RefFilter’s improvements, we compared its performance against two random baselines: an uniform random classifier (coin flip) and a calibrated random classifier that mimics the empirical interference rate (28\%). Using Monte Carlo simulations with 10,000 iterations, we obtained empirical p-values for both F1-score and accuracy. Table~\ref{tab:pvalues} summarizes the empirical p-values.
\begin{table}[h]
\centering
\caption{Empirical p-values against Random Classifiers (Experiment 1)}
\label{tab:pvalues}
\begin{tabular}{lcc}
\hline
\textbf{Comparison} & \textbf{F1-score p} & \textbf{Accuracy p} \\
\hline
RefFilter vs Random (coin-flip) & 0.0981 & 0.0044 \\
RefFilter vs Random calibrated & 0.0222 & 0.2106 \\
\hline
\end{tabular}
\end{table}

RefFilter outperformed both baselines. Compared to the calibrated classifier, RefFilter achieved significantly higher F1-score (p = 0.0222), demonstrating better balance between precision and recall. Furthermore, RefFilter also showed significantly higher accuracy than the coin-flip classifier (p = 0.0096), and a marginally non-significant advantage in F1-score (p = 0.0981). The lack of statistical significance in accuracy against the calibrated random classifier (p = 0.2106) is expected due to class imbalance, as the accuracy is heavily dominated by the large number of true negatives. In other words, even a naive classifier that captures class imbalance can achieve a deceptively high accuracy without actually providing meaningful detection capability. Consequently, F1-score remains a more robust and informative metric to evaluate classifier performance in this scenario.

\todo[inline]{Essa discuss\~{a}o \'{e} super interessante. O Galileu poderia usar algo parecido. Novamente, acho super importante, mas tem uma perspectiva um pouco defensiva. Seria legal a gente discutir com Paulo se seria melhor manter aqui ou em Discussion.}

Moreover, we emphasize that comparison against a random classifier, although common in binary classification benchmarking, does not fully capture the value of our approach. A random method merely answers whether a conflict exists or not, but does not provide actionable information on where conflicts occur (files, classes, or lines involved). In contrast, both pure static analysis and RefFilter provide detailed reports identifying the exact classes, methods, or lines involved in the interference, offering precise localization to assist developers during merge conflict resolution. If a random classifier were required to also guess the location of the interference, its success rate would be close to zero. Thus, even pure static analysis provides actionable and localized information that is practically valuable.


\subsection{Scaling and Diversity Evaluation}
\label{sec:scaling}

\todo[inline]{Continuo achando que essa avalia\c c\~{a}o deveria estar relacionada a uma quest\~{a}o de pesquisa. Algo super simples de ser resolvido. Talvez ``To what extent can RefDiff eliminate potentially false-positive semantic conflicts reported by lightweight static analysis in the wild?''}
\vl{entendi. pode ser mesmo. mas, pessoalmente, ainda penso que esse experimento não responde uma nova pergunta central, mas apenas valida as respostas das RQs 1 e 2 num contexto mais amplo. de toda forma, incluí a sugestão de RQ3 na introdução do experimento, como você sugeriu. Assim, fica mais claro que o experimento 2 complementa o 1 e reforça as respostas das RQs 1 e 2.}

To evaluate the scalability, robustness, and practical relevance of our approach, we conducted a second experiment using a large-scale and diversified dataset comprising 1,087 real-world merge commits drawn from popular open-source projects. Unlike the ground-truth-based setting of Experiment~1, this dataset reflects the variability, noise, and complexity typically found in industrial software development.

\begin{figure}[ht]
    \centering
    \includegraphics[width=0.85\linewidth]{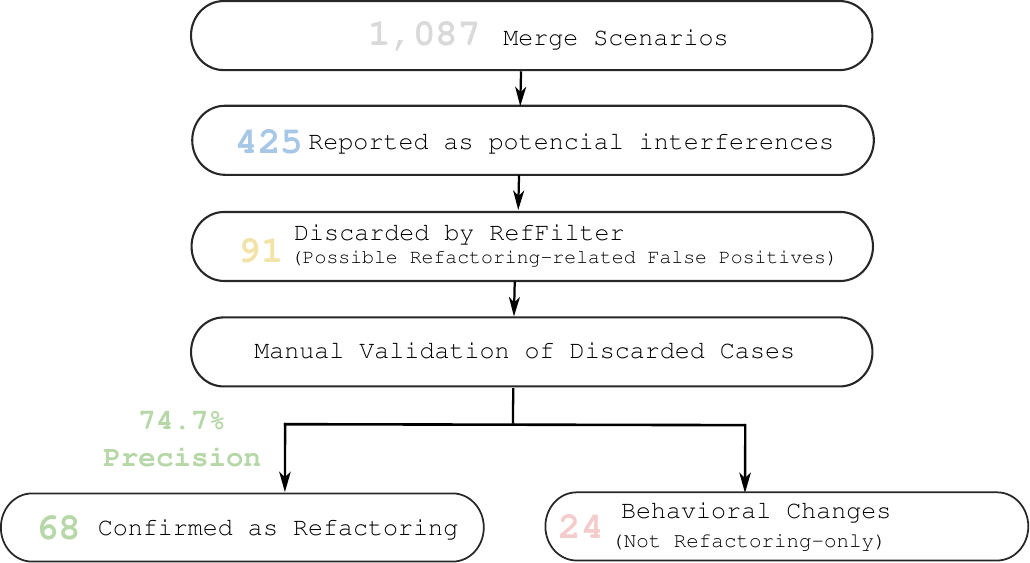}
    \caption{Scaling and diversity results.}
    \label{fig:results2}
\end{figure}

Figure~\ref{fig:results2} presents the results. The baseline static analysis (without any filtering mechanism) reported potential semantic interference in 425 out of the 1,087 merges--- representing 39.1\% of all scenarios. This raw detection rate underscores the power of static analysis to identify candidate conflicts, but also highlights the risk of overreporting due to refactorings.

When RefFilter was applied, 91 of these 425 flagged cases were automatically discarded as likely refactoring-induced false positives--- eliminating 21.4\% of the initial interference reports. \rb{aqui n\~{a}o estamos vendendo muito bem. temos que valorizar a valida\c c\~{a}o destes casos, que foi feita uma amostra estatisticamente significativa e que ABC casos foram validaos. Esse ponto deveria ser valorizado, e relacionado a mais uma ou duas quest\~{o}es de pesquisa discutindo a acur\'{a}cia das ferramentas de detec\ c\~{a}o de refactoring e as causas frequentes de inacur\'{a}cia.}\vl{adicionei um texto no final dese paragrafo para vender que os resultados apontam para boa efetividade do filtro. Também relacionei com o impacto na acurácia das ferramentas de refactoring detection, sem abrir nova RQ}Crucially, manual validation of these discarded cases revealed that 68 were genuine refactorings with no semantic interference, confirming that RefFilter correctly discarded them. The resulting precision of 74.7\% in filtering represents a significant advance: nearly three in every four discarded cases were indeed irrelevant noise successfully filtered out. Although overall accuracy cannot be computed without a complete ground truth, these findings suggest that RefFilter is highly effective at identifying and discarding false positives in large-scale, industrial scenarios, and that current refactoring detection tools (which underpin RefFilter) perform well in practical conditions, though some inaccuracies persist.

\pb{O resumo e a introdução falam apenas na redução de 32\% de FPs, mas isso é só com o dataset pequeno, né? Aqui a taxa seria de no máximo 21\%? Acho que pega mal não dizer isso lá no resumo e introdução. Deveria dizer algo como ``RefFilter reduces at least X\% of FPs, going up to 32\% in one of the analyzed samples.''.}

This outcome is particularly meaningful considering the scale and heterogeneity of the dataset. The tool maintained high effectiveness even in the presence of varied coding styles, project domains, and commit structures--- suggesting strong generalizability and resilience.

The analysis of misclassifications provides important insight. The refactoring type \textit{Change Variable Type} emerged as a major source of filtering errors, appearing in 17 of the 24 incorrect discards \rb{apresentar um exemplo, para valorizar as contribui\c c\~{o}es da pesquisa. esse tipo de resultado poderia aparecer entre os ``bullets'' das contribui\c c\~{o}es na introdu\c c\~{a}o.}\vl{verdade. eu nao expandi essa linha de discussao aqui pela limitacao do espaço mesmo. ja estavamos usando ate ultima linha da pagina 10 antes da revisao}. Conversely, in cases involving simpler structural refactorings such as \textit{Replace Generic With Diamond}, the tool performed exceptionally well--- correctly discarding 11 such scenarios and making only one mistake. This highlights that RefFilter is particularly accurate in identifying and excluding low-risk refactorings, and that further refinements could target more subtle semantic-affecting changes.

Moreover, the decomposition of the detection effort showed that the combined use of RefactoringMiner and ReExtractorPlus was key to achieving robust coverage. Among the 91 discarded scenarios, 29 were identified by RefactoringMiner alone, 7 exclusively by ReExtractorPlus, and 44 required the complementary insights of both tools--- confirming the synergy of using heterogeneous detectors. \rb{essa eh outra contribui\c c\~{a}o \textbf{super interessante}. n\~{a}o sei se trabalhos anteriores recomendam combinar essas duas ferramentas.} \vl{verdade. nao recomendam. o foco da reextractorplus é se vender como superior à refactoringminer, mas vimos que na pratica elas sao melhoes se usadas de forma complementar}

Finally, and most importantly, these results are consistent with those from the labeled dataset in Experiment~1, where 21.2\% of the scenarios were labeled as false positives (i.e., no interference) due to refactorings. In this second, more diverse experiment, we observe a similar trend. Since this is not a labeled dataset, the actual proportion of true and false positives is unknown. Nevertheless, at least 16.0\% of the reported interferences (i.e., the 68 manually validated cases) were correctly filtered as false positives, representing the minimum observed reduction. In the best case, if these 68 are the only false positives among the 425 reported cases, RefFilter would have achieved 100\% precision in false positive filtering.


Finally, this experiment demonstrates not only that our method scales, but also that it retains high precision and remains aligned with ground-truth-based expectations--- all while substantially reducing the noise that hinders practical adoption of static interference detection tools.


\subsection{Qualitative Analysis of Filtering Decisions}
\label{sec:qualitative-analysis}

To complement the quantitative results presented in Sections~\ref{sec:exp1results} and~\ref{sec:scaling}, we conducted a qualitative analysis of representative scenarios in which \emph{RefFilter} either failed to identify a false positive caused by refactoring or incorrectly discarded an interference that should have been reported. These analyses reveal the practical limitations of existing refactoring detection tools and delineate the scope of our technique.

\vspace{0.2cm}
\noindent\textbf{False positives not detected by RefFilter.}  
To analyze cases in which false positives were not successfully filtered, we examined the scenarios from Experiment~1, since this dataset includes ground truth labels. The main sources of missed filtering were the following:

\textit{(1) Multi-step refactorings performed across commits.}  
In some cases, refactorings spanned multiple commits, with only part of the transformation occurring in the target commit. Since refactoring detection tools operate per commit and do not track historical context, such partial transformations were not classified as refactoring. For instance, in one scenario, an partial \emph{encapsulate field} operation replaced the direct use of a field with a call to an accessor method that had been defined in a previous commit:
\begin{lstlisting}[language=Java, caption={Before: Direct access to field}, label=lst:encapsulate-before]
metricRegistry.register("jvm.gc", new GarbageCollectorMetricSet());
\end{lstlisting}

\begin{lstlisting}[language=Java, caption={After: Use of accessor method}, label=lst:encapsulate-after]
getMetricRegistry().register("jvm.gc", new GarbageCollectorMetricSet());
\end{lstlisting}

\textit{(2) Nested and indirect refactorings.}  
Some transformations involved nested operations that semantic tools struggled to capture. One example involved replacing a string literal with a dynamically resolved configuration constant, declared across multiple layers of indirection:

\begin{lstlisting}[language=Java, caption={Before: Hardcoded string literal}, label=lst:nested-before]
.put("node.local", true)
\end{lstlisting}

\begin{lstlisting}[language=Java, caption={After: Resolved constant via getter}, label=lst:nested-after]
//new line
.put(Node.NODE_LOCAL_SETTING.getKey(), true)
//Node class constant declaration
public static final Setting<Boolean> NODE_LOCAL_SETTING = Setting.boolSetting("node.local", false, false, Setting.Scope.CLUSTER);
//constructor in class Setting
public Setting(String key, Setting<T> fallBackSetting, Function<String, T> parser, boolean dynamic, Scope scope)
//getter used in refactor
public final String getKey() {
        return key;
}
\end{lstlisting}

\todo[inline]{Talvez indicar quantos casos se encaixam nas situa\c c\~{o}es (1) e (2) \ldots}\vl{podemos fazer isso futuramente. nao tenho isso mapeado para todos os cenarios}

\vspace{0.2cm}
\noindent\textbf{Non-refactoring cases incorrectly discarded.}  
To investigate cases where potential interference was mistakenly filtered out, we focused on the 24 misclassified scenarios in Experiment~2. Manual inspection revealed three main patterns:
\textit{(1) ``Refactorings'' potentially impacting semantics.}  
Seventeen of the misclassified scenarios involved \texttt{Change Attribute Type}, a refactoring type for which RefactoringMiner claims 100\% precision~\cite{Tsantalis2022}. However, in our manual review, the new types were not always semantically equivalent. For instance, the following change replaces a class from a widely used JSON library with a class from a different API:\pb{completar com ``with different behavior''? porque poderia ser de outra API e ainda assim ser semantically equivalent.}\vl{nós não validamos se o comportamento é de fato diferente}
\begin{lstlisting}[language=Java, caption={Before: Use of Gson library}, label=lst:type-change-before]
import com.google.gson.JsonElement;
(...)
for (Entry<String, JsonElement> e : memb)
\end{lstlisting}

\begin{lstlisting}[language=Java, caption={After: Use of different JSON wrapper}, label=lst:type-change-after]
import foodev.jsondiff.jsonwrap.JzonElement;
(...)
for (Entry<String, JzonElement> e : memb)
\end{lstlisting}

\textit{(2) Minor edits not qualifying as refactorings.}  
Some filtered cases consisted of minor changes unlikely to cause interference, but that do not qualify as refactorings under current definitions. For example, in the change below, the logging message was altered slightly: \pb{esquisito que as ferramentas discardem isso. eles classificam como que refactoring? qual o nome?}\vl{vou verificar o nome. mas reportam sim.}

\begin{lstlisting}[language=Java, caption={Before: Log message}, label=lst:log-before]
getLog().info("skip " + artifact);
\end{lstlisting}

\begin{lstlisting}[language=Java, caption={After: Log message with additional word}, label=lst:log-after]
getLog().info("skip optional " + artifact);
\end{lstlisting}

\textit{(3) Combined refactorings and behavior changes.}  
Some cases involve in the same line or text area behavior-changing edits and  refactorings. For example, in one case, an \texttt{else} block is added at the end of an extracted method, and the refactoring tools labeled all the method lines as refactorings. Although part of the change mirrors a typical refactoring (e.g., method extraction or restructuring), the behavioral addition means the change should not have been filtered.


\todo[inline]{Talvez fechar essa se\c c\~{a}o com um resumo dos findings. Algo como, ``In summary, out of the ABC scenarios identified as refactorings,
  XYZ\% were confirmed through manual analysis. For those classified as false positives, the main causes were \ldots''}
All replication packages, including the datasets, refactoring classification outputs, and detailed experimental results, are available at our online appendix~\cite{online-appendix}. \pb{Melhor assim: are available at our online appendix~\cite{online-appendix}.}
\vl{verdade. mas acho que o problema é o espaço mesmo...}

%% file: sections/discussion.tex
\section{Discussion}
\label{sec:discussion}

The experimental results presented in Section~\ref{sec:results} offer valuable insights into the effectiveness and practicality of refactoring-aware semantic interference detection. In particular, they demonstrate that \emph{RefFilter} can substantially reduce the number of false positives reported by static analysis tools, without introducing a significant loss in recall. This balance between improving precision and maintaining high detection rates is critical for making semantic conflict detection techniques practical in collaborative development settings.

The first experiment, conducted on a benchmark dataset with ground truth labels, showed that \emph{RefFilter} significantly increases precision, with only a minor decrease in recall. Importantly, statistical testing confirmed that the reduction in true positives was not statistically significant, reinforcing that the filtering mechanism does not compromise the utility of the base detection technique.

The second experiment, performed on a large and diverse dataset of over 1,000 merge scenarios, further evaluated the scalability and representativeness of the approach. By manually inspecting a statistically representative sample of discarded interferences, we estimated that approximately 74.7\% of them were indeed false positives—-- confirming that the filter effectively suppresses behavior-preserving changes.

\todo[inline]{Em vez de formularmos as quest\~{o}es como ``does\ldots'', poder\'{i}amos reescrever para algo como: ``To what extent does refactoring-aware semantic conflict detection help reduce false positives?'' A resposta eh bem pr\'{o}xima da atual, remover\'{i}amos apenas o ``yes''.}
\vl{fiz isso!}

\vspace{0.2cm}
\noindent\textbf{RQ1. To what extent does RefFilter reduce false positives compared to traditional static analysis techniques?}

Across both experiments, \emph{RefFilter} consistently reduced the number of false positives. In the labeled dataset, the reduction was 31.25\%, translating into a significative gain in precision. In the larger dataset, out of the 91 scenarios discarded by the filter, manual inspection confirmed that 68 were indeed false positives (i.e., harmless structural changes), demonstrating a strong precision rate for the filtering step. As previously mentioned, since this is not a labeled dataset, the actual proportion of true and false positives is unknown, and at least 16.0\% of the reported interferences (i.e., the 68 manually validated cases) were correctly filtered as false positives. These findings support the hypothesis that many interferences previously reported by static detectors are caused by behavior-preserving refactorings, and can thus be safely ignored. \pb{bom local para trazer aquele percentual minimo de reducao de fps.}

\vspace{0.2cm}
\noindent\textbf{RQ2. To what extent does the reduction in false positives affect the number of false negatives?}
In the benchmark experiment, recall decreased by only 2.5 percentage points, a variation that was shown to be statistically insignificant via McNemar's test. This suggests that the cost of increased precision comes with minimal impact on recall. Additionally, in the large-scale experiment, 24 \pb{de quantas? importante para dar uma ideia do efeito} of the 91 filtered interferences could not be clearly classified as false positives—-- either due to mixed refactorings and business logic or ambiguity in the commit intent. These represent borderline cases where the benefit of manual inspection remains debatable, and do not undermine the practical effectiveness of the filter.
Overall, the results confirm that the filtering approach adopted by \emph{RefFilter} offers a favorable trade-off: it significantly reduces false positives—-- often the most problematic class of error in static interference detection—-- while maintaining a high level of recall.

\vspace{0.2cm}
\noindent\textbf{RQ3. To what extent do the improvements achieved by RefFilter generalize to large-scale, diverse merge scenarios?}
The results from Experiment 2 demonstrate that the improvements observed in the benchmark dataset are generalizable and scalable to industrial settings. RefFilter was applied to a diverse and representative dataset of 1,087 merge scenarios across multiple projects. Despite the absence of ground truth in this dataset, a statistically representative sample of 91 filtered cases was manually validated, revealing that 68 (or 74.7\%) were indeed false positives. This confirms that RefFilter remains effective at suppressing irrelevant interference reports even in heterogeneous and large-scale contexts. The consistency of the findings across both experiments supports the robustness of the approach in practice.

%% file: sections/threats-to-validity.tex
\section{Threats to Validity}
\label{sec:threats}

\textbf{Internal validity.} A potential threat stems from the accuracy of the refactoring detection tools used. Although we rely on state-of-the-art tools, their precision and recall are not perfect. Incorrect or missing refactoring detections may lead to misclassification of interferences. However, our analysis includes complementary tools to mitigate this issue, and the validation results suggest that misclassifications were minimal.

\textbf{Construct validity.} The identification of false positives and true interferences in the large-scale experiment relies on manual inspection of a representative sample. Although we adopted two reviewers at least for each scenario, human judgment introduces inherent subjectivity.

\textbf{External validity.} The benchmark dataset used in Experiment 1 was taken from prior studies, and may not reflect the full diversity of modern development practices. To address this, we constructed a large and unbiased dataset with over 1,000 merge scenarios covering a variety of projects, domains, and commit structures. Still, this is restricted to Java, and generalization to industrial-scale systems may require further validation.

\textbf{Conclusion validity.} Our conclusions about statistical significance are based on McNemar’s test applied to true/false positive changes across methods. While appropriate for the paired nature of the data, small sample sizes in some conditions may limit its power.

%% file: sections/conclusion.tex
\section{Conclusion and Future Work}
\label{sec:conclusion}

This paper presents \emph{RefFilter}, a refactoring-aware tool for static semantic interference detection. Built as a post-analysis layer on top of lightweight static detectors, \emph{RefFilter} aims to reduce false positives by filtering out incorrectly reported interference, which are often caused by changes involving behavior-preserving refactorings. Our formal model defines the conditions under which a potential interference can be safely discarded, and our implementation leverages two complementary refactoring detection tools to perform this filtering in practice.

We evaluate \emph{RefFilter} using both an existing benchmark dataset with ground truth and a newly constructed dataset of 1,087 diverse merge scenarios. Results show that  \emph{RefFilter} reduces the number of false positives by nearly 32\% on the labeled dataset, with a non
significant increase in false negatives.\pb{melhor talvez focar aqui nos numeros de reducao de fp, nao precision. alinhar com resumo e intro. ou falar nos dois, para evitar que o leitor ache que foi um erro. pode-se ler precision e pensar em FP porque boa parte do texto focou em FP.} These findings indicate that refactorings are a major source of noise in static detection pipelines and that they can be effectively mitigated with lightweight techniques.

In addition to empirical evidence, we introduce a formal characterization of refactoring-induced false positives and demonstrate its practical application in industrial scenarios. We also identify key cases where refactoring detection tools fail, either due to multi-step transformations, complex nesting, or semantic ambiguity, outlining directions for future improvement.

